# Holistic random encoding for imaging through multimode fibers

Hwanchol Jang,[1] Changhyeong Yoon,[2] Euiheon Chung,[3] Wonshik Choi,[2] and Heung-No Lee[1,*]

[1] *School of Information and Communications, Gwangju Institute of Science and Technology, Gwangju 500-712, South Korea*
[2] *Department of Physics, Korea University, Seoul 136-701, South Korea*
[3] *Department of Medical System Engineering and School of Mechatronics, Gwangju Institute of Science and Technology, Gwangju 500-712, South Korea*
[*]*heungno@gist.ac.kr*

**Abstract:** The input numerical aperture (NA) of multimode fiber (MMF) can be effectively increased by placing turbid media at the input end of the MMF. This provides the potential for high-resolution imaging through the MMF. While the input NA is increased, the number of propagation modes in the MMF and hence the output NA remains the same. This makes the image reconstruction process underdetermined and may limit the quality of the image reconstruction. In this paper, we aim to improve the signal to noise ratio (SNR) of the image reconstruction in imaging through MMF. We notice that turbid media placed in the input of the MMF transforms the incoming waves into a better format for information transmission and information extraction. We call this transformation as holistic random (HR) encoding of turbid media. By exploiting the HR encoding, we make a considerable improvement on the SNR of the image reconstruction. For efficient utilization of the HR encoding, we employ sparse representation (SR), a relatively new signal reconstruction framework when it is provided with a HR encoded signal. This study shows for the first time to our knowledge the benefit of utilizing the HR encoding of turbid media for recovery in the optically underdetermined systems where the output NA of it is smaller than the input NA for imaging through MMF.

## 1. Introduction

Multimode fibers (MMF) support *multimode* propagation of light such that the light travels not only along the cylindrical axis of the core, *single* mode if it does, but also along multiple different paths with non-zero components in traverse directions.

The use of MMF for imaging has drawn great interests recently [1–7]. Current endoscopic imaging systems in clinics are based on bundles of fibers where each fiber transfers the signal corresponding to a single pixel of the final image. The multiple propagation characteristic of MMF allows a complex image to be transferred through not a bundle of fibers but only with a single fiber. This enables the miniaturization of imaging systems. Thus, MMF is expected to become a significantly important part for minimally invasive endoscopic imaging where a fiber with needle-like dimensions can transfer complex images. However, there is an intrinsic limitation in imaging through MMF on the spatial image resolution which is imposed by the low numerical aperture (NA) of available MMF [4–7]; the typical NA of MMFs with large number of propagation modes range from 0.2 to 0.5 [5] whereas the NA of optical lenses reaches up to 0.95.

It has been demonstrated that the problem of the low resolution given by the low NA of MMFs can be relaxed by the use of turbid media in conjunction with MMFs [6,7]. Wave propagation through turbid media, such as white paint, ground glass, and biological tissue, produces complex speckle patterns in the image plane due to multiple scattering of waves in the media. Multiple scattering of waves, referring to the phenomenon where the light waves are forced to deviate from a straight trajectory due to refractive index inhomogeneity through which they pass. It is obvious that this multiple scattering process hinders accurate transferring of images through turbid media. Recently, interesting results were reported that the multiple scattering process in turbid media can be used for overcoming the resolution limit determined by the NA of the optical systems [8,9]. The NA of an optical system sets the maximum incident angle $\theta_{max}$ of the incoming waves that can be accepted by the system. In wide-field imaging, the multiple scattering through turbid media changes the directions of input waves and some of the waves with large incident angles beyond the acceptance angle of a usual lens can be redirected to the detector [8]. Thus, the effective input NA of the lens becomes increased. In point scanning imaging, multiple scattering is combined with wave-front shaping and makes it possible to focus the light beam to a point smaller than the diffraction limit given by the NA [9]. These resolution improvements by turbid media can be made with the same principle in imaging through MMF when it is used with turbid media. It was found that waves with larger incident angles than the acceptance angle of the MMF can be transferred in the wide-field imaging [7], and a smaller focusing point can be made in point scanning imaging [6].

Here, we show that the redirection of waves is not the only useful characteristic of turbid media. Multiple scattering in turbid media scrambles the waves in a seemingly randomized way and this brings forth other positive effects as well. Random scrambling of different modes of waves converts the object wave into a speckle pattern. The object cannot be directly observed looking at the speckle pattern with bare eyes, and thus some form of an inverse operation is necessary such as descrambling for imaging or wave-front shaping for focusing [8–12]. Thus, random scrambling gives an impression that it is only a hindrance; on the contrary, this traditional perspective can be challenged and improved.

Waves scattered from an object are made of the superposition of multiple different modes. Those modes travel through the turbid medium and become scrambled. We note that each mode of the scrambled waves in fact contains *holistic* information of the object. A signal mode is *holistic* in the sense that the mode contains the information of the whole modes of the incoming waves. Each mode of incoming waves propagates though the turbid medium is scattered and redirected into many different modes at the output of the turbid media. One mode is scattered into almost all the modes of the output waves. This is to mean that a single output mode is made out of the superposition of many if not all incoming waves. In addition, each mode of the incoming waves went through a propagation path completely different from other modes. Thus, each output mode offers an independent view of the same object. In this sense, the random scrambling can be viewed as a beneficial encoding process, which provides multiple independent outlooks of the whole waves from an object. We will refer to this signal transforming process in turbid media as *holistic random* (HR) encoding.

In this paper, we aim to make an efficient use of the HR encoding of the multiple scattering in turbid media and improve the image quality in wide-field imaging through MMF. We consider an imaging system through MMF where a turbid medium is placed at the input end of the MMF. In this case, the imaging system becomes underdetermined because the output NA of the MMF is smaller than the input NA of it. The information of object waves is transferred to a less number of wave modes than that of the input wave modes. Considering that the degree of freedom of a signal is reduced, it is not easy to transmit information of the object waves without loss. To compound the matter, the recovery of the object waves is not easy as well since the dimension (the number of elements) of the observed signal is smaller than that of the original signal to be estimated. Here, we show that the HR encoding of the turbid media enables much improved information transmission and signal reconstruction. We employ sparse representation (SR) framework [13–21] and show that the object information can be extracted at an improved fidelity when the signal is HR encoded. In many literatures, SR has been shown to provide superb estimation of the original signal from a smaller number of measurements than the dimension of the original signal [13–21]. Previously, SR was shown to be beneficial in imaging through turbid media [11,12]. In [11], SR was shown to suppress speckles in reconstructed images. It was shown in [12] that SR recovers the image well in the situation where the number of pixels in the CCD array is smaller than that the pixels of the original image. This paper provides the first result that SR can be used to improve the image reconstruction quality in imaging through MMF when the imaging system is underdetermined due to the use of turbid media at the input end of the MMF.

This paper is organized as follows: Section 2 describes the system for imaging through MMF, and Section 3 investigates the HR encoding by the multiple scattering in turbid media and links it to the SR framework for image reconstruction. Experimental results are discussed in Section 4, and Section 5 concludes the paper.

**2. System description**

*2.1 Experimental set-up*

We consider an imaging system[1] where the object wave propagates through a single MMF with 1 m of length (NA of 0.22; Thorlabs, M14L01) and recorded at the CCD array. The input facet of the MMF is randomly coated by ZnO nanoparticles.

The experimental schematic of the imaging is depicted in Fig. 1. Figure 1 (a) describes the calibration stage where the transmission matrix (TM) of the coated MMF is measured. TM is a collection of responses of the coated MMF to a set of incoming plane waves ($\lambda = 633$ nm) with $N$ different incident angles to the input facet of the coated MMF. We used $N = 4000$ different angles for our TM. For preparation of the plane waves, no object is presented in the object plane. The incident angle of the plane wave is controlled by a galvanometer.

The transferring of an object wave is described in Fig. 1 (b). Once the TM is measured and becomes available, the response of the coated MMF to the object wave is measured at the CCD. The object wave is distorted in the coated MMF due to the multiple scattering in the turbid medium of ZnO nanoparticles and the interference among waves with different propagation modes inside the MMF. For the object, we use a sample similar to USAF target.

Now with the TM and the distorted object image, the object wave is recovered by computation. Here, all the measurements are post processed by the off-axis holography [8] to obtain the E-field images. We use MMF with the NA of 0.22 in the experiment. We fix the NA of the sub-systems followed by the MMF a little bit larger, 0.24, than 0.22. By doing this, we can capture the most of the signal from the MMF even though there are some experimental mismatches, for example, the error in the alignment on the optical axis.

---

[1] The baseline method considered in this paper follows the turbid lens imaging (TLI) system in [8].

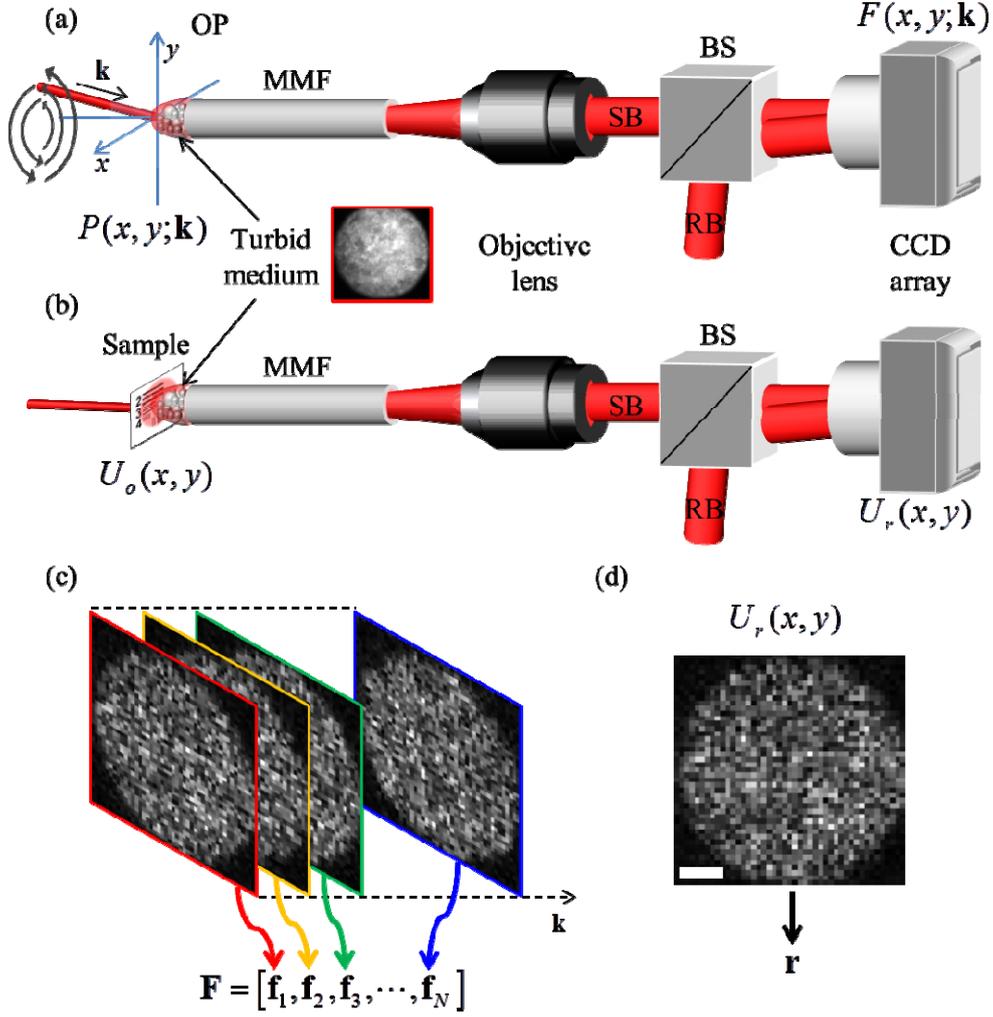

Fig. 1. Experimental schematics of imaging through turbid media coated MMF. (a) Recording of the TM. OP: object plane, MMF: multimode fiber, BS: beam splitter, SB: sample beam, RB: reference beam. The photo in the red box shows the image of the input surface of the turbid media coated MMF. (b) Recording of the distorted object wave. (c) The recorded TM. Only the intensities are shown in the image but the phases are estimated as well; thus the columns are complex valued vectors. (d) The recorded distorted object wave. Again, only the intensity is shown here. Scale bar: 10 μm.

2.2 Image recovery using transmission matrices

The object wave $U_o(x, y)$ is decomposed into a set of plane waves with different propagation directions as follows

$$U_o(x, y) = \sum_{\mathbf{k}} A_o(\mathbf{k}) P(x, y; \mathbf{k}) \quad (1)$$

where $\mathbf{k} := k_x \mathbf{i}_x + k_y \mathbf{i}_y + k_z \mathbf{i}_z$ is the wave vector ( $k_x / 2\pi = \sin\theta_x / \lambda$ , $k_y / 2\pi = \sin\theta_y / \lambda$ , $k_z / 2\pi = \sin\theta_z / \lambda$ , and $(\theta_x, \theta_y, \theta_z)$ is the angle of propagation), $\mathbf{i}_x$, $\mathbf{i}_y$, and $\mathbf{i}_z$ are unit vectors in *x*, *y*, and *z* directions (the optical axis is in the *z* direction), respectively, $P(x, y; \mathbf{k})$ is the plane wave with the propagation direction **k**, and $A_o(\mathbf{k})$ is the angular spectrum of the object wave. The distorted object wave $U_r(x, y)$ after propagation through the coated MMF is expressed as

$$U_r(x, y) = \sum_{\mathbf{k}} A_o(\mathbf{k}) F(x, y; \mathbf{k}) \qquad (2)$$

where $F(x, y; \mathbf{k})$ is the response wave of the coated MMF for a single plane wave $P(x, y; \mathbf{k})$. Now using the vector notations, the distorted object wave can be expressed as follows

$$\mathbf{r} = \mathbf{F}\mathbf{a} \qquad (3)$$

where $\mathbf{r} \in \mathbb{C}^M$ and $\mathbf{a} \in \mathbb{C}^N$ are the vectorized versions of $U_r(x, y)$ and $A_o(\mathbf{k})$, and $\mathbf{F} \in \mathbb{C}^{M \times N}$ is the TM each column of which is the vectorized version of the $F(x, y; \mathbf{k})$; there are $N$ different propagation directions (modes) **k** considered.

From the distorted object wave, the angular spectrum of the object wave is estimated in [7] by using the pseudo inversion (PINV) method,

$$\hat{\mathbf{a}}_{pinv} = \mathbf{F}^{-1}\mathbf{r} \qquad (4)$$

where $\mathbf{F}^{-1}$ is the PINV matrix of **F**. Using the estimated angular spectrum, the object wave in the object plane can be reconstructed.

### 3. Imaging through MMF in conjunction with turbid media

Recall that the input facet of the MMF is coated by a turbid medium. Due to the multiple scattering process through the turbid medium, some of those waves whose incident angles to the medium are larger than the acceptance angle determined by the NA of the MMF, $\theta_z > \theta_{max}$, are redirected, $\theta_z \to \theta_z'$, and are coupled to the MMF, $\theta_z' \leq \theta_{max}$. This introduces more modes of the object (the green arrows in Fig. 2 (b)) to the detector. The NA of a MMF is proportional to the square root of the number of modes captured by the MMF. As a result, it was shown in experiments [7] that the effective NA of the input side of the MMF increases.

However, we notice that the increase in the effective NA causes the image capturing system to be underdetermined where the input NA is larger than the output NA (Fig. 2). Larger number of modes of the object waves than the available propagation modes in the MMF is captured. This gives a challenge in the recovery of the object. In general, the output signal of an underdetermined system does not convey all the information of the input signal. Besides, even in the cases where there are no information losses, it is also well known from the linear algebra [22] that the correct estimation of a signal in the underdetermined systems is not easy for it has more than one solution. Note that the use of another turbid medium in the output facet does not change the situation because the number of propagation modes inside the MMF is still the same.

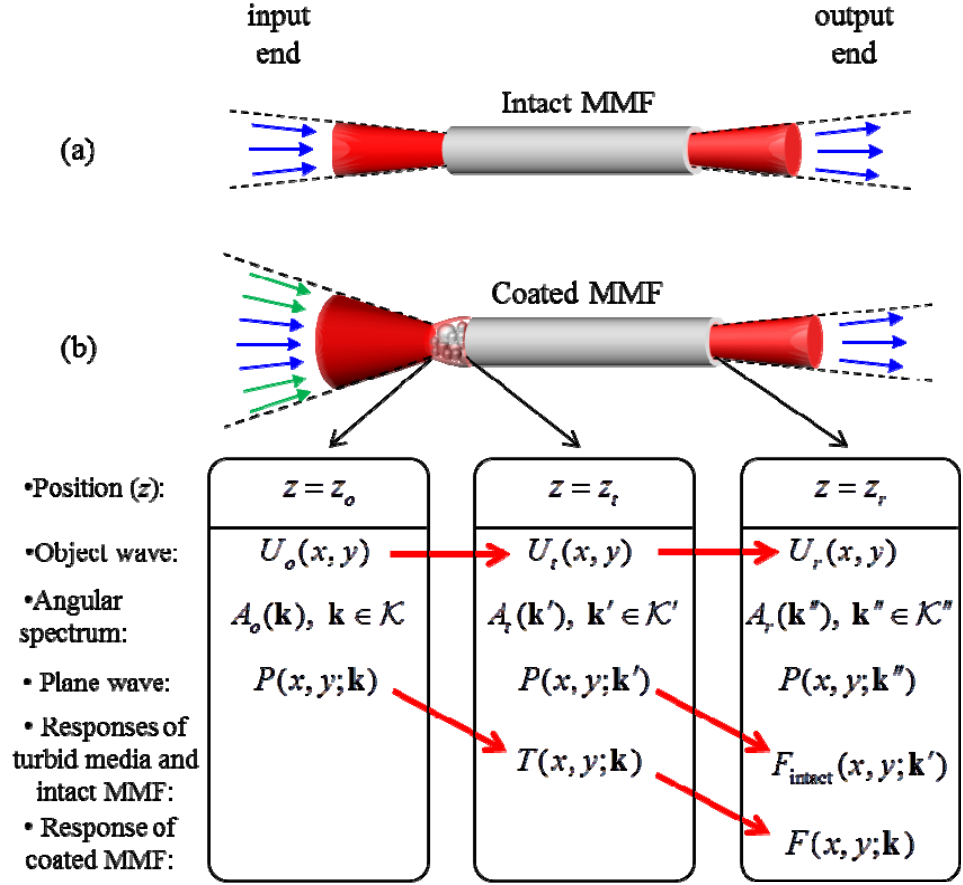

Fig. 2. Acceptance angle, exit angle, and their corresponding modes of waves for (a) intact MMF and (b) coated MMF. In (b), more modes of input waves (green arrows) can be captured in coated MMF. But, the number of modes of output waves is the same as that of intact MMF.

*3.1 HR encoding*
Multiple scattering in turbid media provides HR encoding to the incoming wave. HR encoding gives the two kinds of encoding, holistic encoding and random encoding. 1) Holistic encoding: we define "holistic encoding" to be a signal transforming process which makes each component of the transformed signal to contain all the components of the original signal with a certain format, especially in this paper, a weighted summation of them. Each input mode $P(x, y; \mathbf{k})$ of the incoming waves $U_o(x, y)$ in [Eq. (1)] to the turbid medium experiences multiple scattering. The object wave $U_t(x, y)$ after propagation through the turbid medium (refer to Fig. 2 (b)) is expressed as

$$U_t(x, y) = \sum_{\mathbf{k} \in \mathcal{K}} A_o(\mathbf{k}) T(x, y; \mathbf{k}) \tag{5}$$

where $\mathcal{K}$ is the set of modes $\mathbf{k}$ which propagate through the coated MMF and reach to the detector and $T(x, y; \mathbf{k})$ is the response wave of the turbid medium to $P(x, y; \mathbf{k})$. The response waves $T(x, y; \mathbf{k})$ at the output of the turbid medium for input waves $P(x, y; \mathbf{k})$ with a single mode $\mathbf{k}$ has many modes of waves. This is because the multiple scattering process in the turbid medium changes the directions of the waves in a randomized manner. $T(x, y; \mathbf{k})$ is expressed as

$$T(x, y; \mathbf{k}) = \sum_{\mathbf{k}' \in \mathcal{K}'} t(\mathbf{k}'; \mathbf{k}) P(x, y; \mathbf{k}') \tag{6}$$

where $\mathbf{k}'$ is the wave vector after propagation through the turbid medium, $\mathcal{K}'$ is the set of modes $\mathbf{k}'$ which propagate through the MMF and reach to the detector, $t(\mathbf{k}'; \mathbf{k}) \in \mathbb{C}$ is the contribution of a mode $\mathbf{k}$ of input waves to a mode $\mathbf{k}'$ of output waves, and $P(x, y; \mathbf{k}')$ is the plane wave with the propagation direction $\mathbf{k}'$.

Here, $t(\mathbf{k}'; \mathbf{k})$ is well approximated by independent and identically distributed (i.i.d.) complex valued Gaussian random variable. It was found in [23,24] that the output waves of a turbid medium at a spatial plane $(x', y')$ when a mode of waves is transmitted through the medium, $t(x', y'; \mathbf{k})$, are i.i.d. complex valued Gaussian random variables provided that the number of independent scatters is large; $t(x', y'; \mathbf{k})$ and $t(\mathbf{k}'; \mathbf{k})$ are a two-dimensional Fourier transform pair. This has been also supported in the experiments [12,25]. The distribution of the eigenvalues of the TM composed of $t(x', y'; \mathbf{k})$ was close to that of i.i.d. Gaussians [25]. The coherence ($\mu_0$ in Sec 3.2) of the TM of $t(x', y'; \mathbf{k})$ behaved similarly to that of i.i.d. Gaussians [12]. We know that the Fourier transform of a Gaussian random matrix is another Gaussian random matrix. Thus, the contribution $t(\mathbf{k}'; \mathbf{k})$ follows i.i.d. complex valued Gaussian, too.

Using [Eq. (6)], the object wave $U_t(x, y)$ in [Eq. (5)] can be expressed as

$$U_t(x, y) = \sum_{\mathbf{k}' \in \mathcal{K}'} A_t(\mathbf{k}') P(x, y; \mathbf{k}') \tag{7}$$

where the angular spectrum $A_t(\mathbf{k}')$ of $P(x, y; \mathbf{k}')$ is

$$A_t(\mathbf{k}') = \sum_{\mathbf{k} \in \mathcal{K}} t(\mathbf{k}'; \mathbf{k}) A_o(\mathbf{k}). \tag{8}$$

Here, the probability that $t(\mathbf{k}'; \mathbf{k}) = 0$ is very small; the probability that a realization of complex Gaussian random variable is equal to zero approaches zero. The angular spectrum $A_t(\mathbf{k}')$ of a mode $\mathbf{k}'$ of the output waves is now the combination of the angular spectrum $A_o(\mathbf{k})$ of all the modes $\mathbf{k}$. Thus, a mode of the output waves contains holistic information of the input waves. 2) Random encoding: we define "random encoding" to be a multiplexing process in which the contributions of a component of the original signal to components of the output signal are realizations of random variables which are independent. As it was discussed in the previous paragraph, $t(\mathbf{k}'; \mathbf{k})$ is approximated by an independent random variable. Thus, the angular spectrum $A_t(\mathbf{k}')$ of each mode $\mathbf{k}'$ of the output waves shows an independent view of the same object $U_o(x, y)$. Now, we expect that the waves $U_o(x, y)$ scattered from an object is HR encoded in the turbid medium.

This HR encoding has a couple of desirable aspects to send information of object waves in underdetermined systems. i) The information of all the input modes is transferred no matter how few modes are in the output waves, if it is more than one. It is because each output mode captures information about all the input modes of the object wave. ii) It sends information of the object waves in an efficient manner. The information of a mode is not redundant to that of other modes as each output mode captures unique information about the object wave. Due to these two aspects, the HR encoded signal is expected to be appropriate for information transmission in underdetermined systems. In the literature, it is shown that HR encoded signals can transmit enough information for a certain kind of signals in underdetermined systems [16–18].

Now, let us see the effect of HR encoding on the object wave after propagation through the MMF. The HR encoded waves $U_t(x, y)$ at the output of the turbid medium are propagated through the MMF. The object wave $U_r(x, y)$ after propagation through the MMF (Fig. 2 (b)) is expressed as

$$U_r(x, y) = \sum_{\mathbf{k}' \in \mathcal{K}'} A_t(\mathbf{k}') F_{\text{intact}}(x, y; \mathbf{k}') \tag{9}$$

where $F_{\text{intact}}(x, y; \mathbf{k}')$ is the response wave of the intact MMF to $P(x, y; \mathbf{k}')$. $F_{\text{intact}}(x, y; \mathbf{k}')$ usually has more than one propagation modes of waves in MMF. This is because that the connection of a mode of incoming waves to only a single propagation mode in MMF is almost infeasible. $F_{\text{intact}}(x, y; \mathbf{k}')$ is expressed as

$$F_{\text{intact}}(x, y; \mathbf{k}') = \sum_{\mathbf{k}'' \in \mathcal{K}''} f_{\text{intact}}(\mathbf{k}''; \mathbf{k}') P(x, y; \mathbf{k}'') \tag{10}$$

where $\mathbf{k}''$ is the wave vector after propagation through the MMF, $\mathcal{K}''$ is the set of modes $\mathbf{k}''$ which are captured at the detector, $f_{\text{intact}}(\mathbf{k}''; \mathbf{k}')$ is the contribution of a mode $\mathbf{k}'$ of input waves to a mode $\mathbf{k}''$ of output waves, and $P(x, y; \mathbf{k}'')$ is the plane wave with the propagation direction $\mathbf{k}''$. Using [Eq. (8)], [Eq. (9)], and [Eq. (10)], the object wave $U_r(x, y)$ can be rewritten as

$$U_r(x, y) = \sum_{\mathbf{k}'' \in \mathcal{K}''} A_r(\mathbf{k}'') P(x, y; \mathbf{k}'') \tag{11}$$

where the angular spectrum $A_r(\mathbf{k}'')$ of $P(x, y; \mathbf{k}'')$ is

$$A_r(\mathbf{k}'') = \sum_{\mathbf{k} \in \mathcal{K}} C(\mathbf{k}''; \mathbf{k}) A_o(\mathbf{k}), \tag{12}$$

and the weight $C(\mathbf{k}''; \mathbf{k})$ of $A_o(\mathbf{k})$ for $A_r(\mathbf{k}'')$ is

$$C(\mathbf{k}''; \mathbf{k}) = \sum_{\mathbf{k}' \in \mathcal{K}'} t(\mathbf{k}'; \mathbf{k}) f_{\text{intact}}(\mathbf{k}''; \mathbf{k}'). \tag{13}$$

Please recall that a larger number of modes $\mathbf{k}$ of the incoming waves than that of propagation modes $\mathbf{k}''$ in the intact MMF (equivalently $\mathbf{k}'$) are captured with the help of the turbid medium in the coated MMF, $|\mathcal{K}| > |\mathcal{K}''|$. We see that the angular spectrum $A_o(\mathbf{k})$ for $\mathbf{k} \in \mathcal{K}$ of the object waves $U_o(x, y)$ are transferred through waves $P(x, y; \mathbf{k}'')$ with a smaller number of propagations modes $\mathbf{k}'' \in \mathcal{K}''$ in [Eq. (12)]. Thus, the reconstruction of the $A_o(\mathbf{k})$ (or $U_o(x, y)$) from the $A_r(\mathbf{k}'')$ (or $U_r(x, y)$) is an underdetermined problem.

We see that the signal is still holistically encoded in the output waves $U_r(x, y)$ in [Eq. (12)]. Every component of the angular spectrum $A_o(\mathbf{k})$ of the input waves is contained in each single components of the angular spectrum $A_r(\mathbf{k}'')$ of the output waves. The random encoding property of $U_t(x, y)$ is not inherited to $U_r(x, y)$ since $C(\mathbf{k}'';\mathbf{k})$ is not independent for different $\mathbf{k}''$. But, we will see that the correlation between $C(\mathbf{k}'';\mathbf{k})$ with different $\mathbf{k}''$ is small in Fig. 3 (f). This is desirable for image recovery (Sec 3.2). A further explanation on $C(\mathbf{k}'';\mathbf{k})$ is given in Sec 3.3. Please note that we use $P(x, y;\mathbf{k}'')$ for the modes of the output waves of the MMF and for the propagation modes in the MMF interchangeably without distinction as they are one-to-one mapped.

*3.2 Sparse representation*

SR is a signal representation framework, which has received great interests since it can be used to estimate a signal even in the underdetermined systems [13–21]. HR encoding increases the oversampling ratio of the underdetermined systems in which a signal is estimated correctly [16–18]. It is well known that there are two conditions for successful application of the SR framework for recovery of original signal from its HR encoded one. First, the signal **a** is compressible. A compressible signal means that the signal **a** is well approximated with a small number of nonzero elements in **a**, say $K$ where $K \ll N$. The object signals of interest in this paper are natural signals and we have many research results showing that they are compressible. It is well known that most natural images are well approximated with only a few elements in the Wavelet domain [15]. Not only with the Wavelet domain, if the signal is represented with a few elements in any other orthogonal signal bases, the signal is compressible [16]. Second, the measurement matrix **F** needs to be incoherent. We say a matrix is incoherent if the cross-correlations of columns of the matrix are small. This follows the conventional meaning for incoherence of a matrix in [14,17,20]. Note that this incoherence is different from that in optics which is typically a phase relationship among waves. The incoherence of a measurement matrix can be measured in its Gram matrix $\mathbf{D} = \mathbf{F}^*\mathbf{F}$. We assume that the norm of each column of **F** is normalized to be one. The amplitude $|d_{ij}|$ of the off-diagonal elements of **D** indicates the cross-correlation of different $i^{\text{th}}$ and $j^{\text{th}}$ columns of the measurement matrix where $d_{ij}$ is the $(i,j)^{\text{th}}$ element of **D** and $|\cdot|$ denotes the absolute value of the complex number.

Several different measures are used for incoherence of a measurement matrix. The simplest one is i) the largest off-diagonal element in the Gram matrix, $\mu_0 \triangleq \max_{i,j} |o_{ij}|$ where $o_{ij}$ is the $(i,j)^{\text{th}}$ element of **O**=**D**-**I** [14,17,20]. But, this does not characterize the incoherence of a measurement matrix well for it only considers the most extreme case [20]. The other two measures are ii) the size of the smallest group of off-diagonal elements in a single row of the Gram matrix which have the sum greater than one, $\mu_1 \triangleq \min_i \left( \min |J| \text{ s.t. } \sum_{j \in J} |o_{ij}| \geq 1 \right)$ where $|J|$ is the cardinality of an index set $J$ [14], and iii) the maximum value of the summation of $M$ off-diagonal elements in a single row of the Gram matrix, $\mu_2 \triangleq \max_{|J|=M} \max_{i \notin J} \sum_{j \in J} |o_{ij}|$ [20]. Those two measures provide a little bit more of the general behavior of a measurement matrix. However, they still do not provide an overall behavior of the measurement matrix as they also reflect extreme cases by taking the minimum size or the maximum sum; actually, the two measures are designed to provide theoretical bounds of $M$ which guarantee the successful estimation of sparse signals, which are signals with small numbers of nonzero elements, in the SR framework.

In this paper, we aim to use $\mu_3 \triangleq \left|\left\{o_{ij} \mid \left|o_{ij}\right| > \alpha\right\}\right| / \left(N^2 - N\right)$ the fraction of the off-diagonal elements of a Gram matrix whose absolute values are comparable, $\left|o_{ij}\right| > \alpha$ where $\alpha$ is the degree of comparableness ($0 < \alpha < 1$), to the value of the diagonal elements as the measure of incoherence. The rationale for this is i) that it does not take any extreme values, ii) that it is coherent to the previous three measures, $\mu_0$, $\mu_1$, and $\mu_2$, as more number of large off-diagonal elements is likely to lead to values of the three measures which indicate the matrix is less incoherent, and iii) that the use of fractions is appropriate to compare the incoherence of the two different Gram matrices with different size; this is the case we consider in this paper. It is desirable to have smaller $\mu_3$. Construction of an incoherent measurement matrix in a deterministic fashion for an underdetermined system is known to be not easy. Fortunately, in the literature, it is shown analytically or empirically that many kinds of randomly generated matrices are incoherent [16–19]. That is, we can make good measurement matrices by using random generation, without a careful precise matrix design.

Now we aim to explain how SR framework can be utilized to recover the signal successfully in our turbid lens based MMF imaging. The current state-of-the-art SR systems can recover a signal with $K$ nonzero elements, the so-called $K$-sparse signal, correctly with just $M = O\left(K \log(N/K)\right)$ number of random measurements [17].

The estimation in SR can be done by finding solution of the following problem [17]

$$\hat{\mathbf{a}}_{SR} = \arg\min_{\mathbf{a}} \left\|\mathbf{\Psi}^*\mathbf{a}\right\|_1 \text{ subject to } \mathbf{r} = \mathbf{F}\mathbf{a}, \tag{14}$$

where $\mathbf{\Psi}$ is the sparsifying basis in which the signal $\mathbf{a}$ can be approximated with just a small number of nonzero elements, $(\cdot)^*$ denotes the conjugate transpose of a matrix, and $\left\|\cdot\right\|_1$ denotes the L1 norm, i.e., the sum of the absolute values of the vector elements.

*3.3 Effect of the random scrambling of turbid media on TM*

As it was discussed in Sec 3.1, turbid media provide HR encoding to the object waves. In this subsection, we show that HR encoding provides incoherent TM to MMF. We focus on how improved the coated MMF is compared to the intact MMF in terms of incoherence. As we have discussed in Sec 3.2, we will use $\mu_3$ as the measure of incoherence.

TM is a collection of responses $F(x, y; \mathbf{k})$ of the MMF to a set of incoming plane waves $P(x, y; \mathbf{k})$ with $N$ different modes $\mathbf{k}$ (Sec 2.2). Let us consider first the TM of an intact MMF without a turbid medium deposited. For an intact MMF, the response of the MMF $F(x, y; \mathbf{k})$ is $F_{\text{intact}}(x, y; \mathbf{k})$ in [Eq.(10)]; here, $\mathbf{k}$ and $\mathbf{k}'$ are the same. The TM of the MMF consists of $F_{\text{intact}}(x, y; \mathbf{k})$ with different modes $\mathbf{k}$. As it was told in Section 3.1, $F_{\text{intact}}(x, y; \mathbf{k})$ is a superposition of waves $P(x, y; \mathbf{k}'')$ for more than one propagation modes $\mathbf{k}''$ in MMF. $F_{\text{intact}}(x, y; \mathbf{k})$ in [Eq.(10)] can be written as

$$F_{\text{intact}}(x, y; \mathbf{k}) = \sum_{\mathbf{k}'' \in S_{\text{intact}}(\mathbf{k})} f_{\text{intact}}(\mathbf{k}''; \mathbf{k}) P(x, y; \mathbf{k}'') \tag{15}$$

where $S_{\text{intact}}(\mathbf{k}) := \left\{\mathbf{k}'' \mid f_{\text{intact}}(\mathbf{k}''; \mathbf{k}) \neq 0\right\}$ is the set of excited propagation modes in the MMF when plane waves $P(x, y; \mathbf{k})$ with the mode $\mathbf{k}$ are inserted; $S_{\text{intact}}(\mathbf{k}) \subset \mathcal{K}''$. We found that a small number of propagation modes in the MMF are excited when waves with a single mode are inserted (Fig. 3 (c)); this is serious especially when the incident angles $\theta_z$ of the incoming waves are small. With a small number of propagation modes $P(x, y; \mathbf{k}'')$, there are not many incoherent $F_{\text{intact}}(x, y; \mathbf{k})$ available because they are the combinations of the few plane waves whose modes $\mathbf{k}''$ are included in $S_{\text{intact}}(\mathbf{k})$; it is analogous to generating incoherent vectors which are linear combinations of a small number of basis columns. Thus, it is not easy to fill up the columns of the TM with incoherent $F_{\text{intact}}(x, y; \mathbf{k})$ for $\mathbf{k}$ with small incident angles. This results in a not incoherent TM.

We show in Fig. 3 (a) some of the $F_{\text{intact}}(x, y; \mathbf{k})$ with several $\mathbf{k}$, respectively. We consider discrete incident angles $\mathbf{k}_i$ for $1 \le i \le 2000$. Among the responses $F_{\text{intact}}(x, y; \mathbf{k}_i)$ of all the considered incident angles $\mathbf{k}_i$, we show $\mathbf{k}_i$ with $i$=1, 101, 201, and 1001. The incident angle changes with spiral shape starting from the center. A small index $i$ of the response means that it is for that with a small incident angle $\theta_z$. We see that the responses $F_{\text{intact}}(x, y; \mathbf{k}_i)$ of smaller incident angles ($i$=1, 101, 201) are not complex due to lack of the available propagation modes $P(x, y; \mathbf{k}'')$ (Fig. 3 (c)). Here, we mean by complex that the value in each pixel $(x, y)$ of the response $F_{\text{intact}}(x, y; \mathbf{k}_i)$ changes enough from those in its neighboring pixels. This can be seen in the autocorrelation of $F_{\text{intact}}(x, y; \mathbf{k}_i)$,

$$R_{FF;\text{intact}}(x, y; \mathbf{k}_i) \triangleq \left| \sum_{x'} \sum_{y'} F_{\text{intact}}(x', y'; \mathbf{k}_i) F_{\text{intact}}^*(x' + x, y' + y; \mathbf{k}_i) \right|$$

in Fig. 3 (b). It looks complex around $i$=1001. But it is also made of plane waves with only several different modes, not all the modes considered. Compared to $R_{FF}(x, y; \mathbf{k}_i)$ in Fig. 3 (e), $R_{FF;\text{intact}}(x, y; \mathbf{k}_i)$ has more non-ignorable values at $x \ne 0$ or $y \ne 0$.

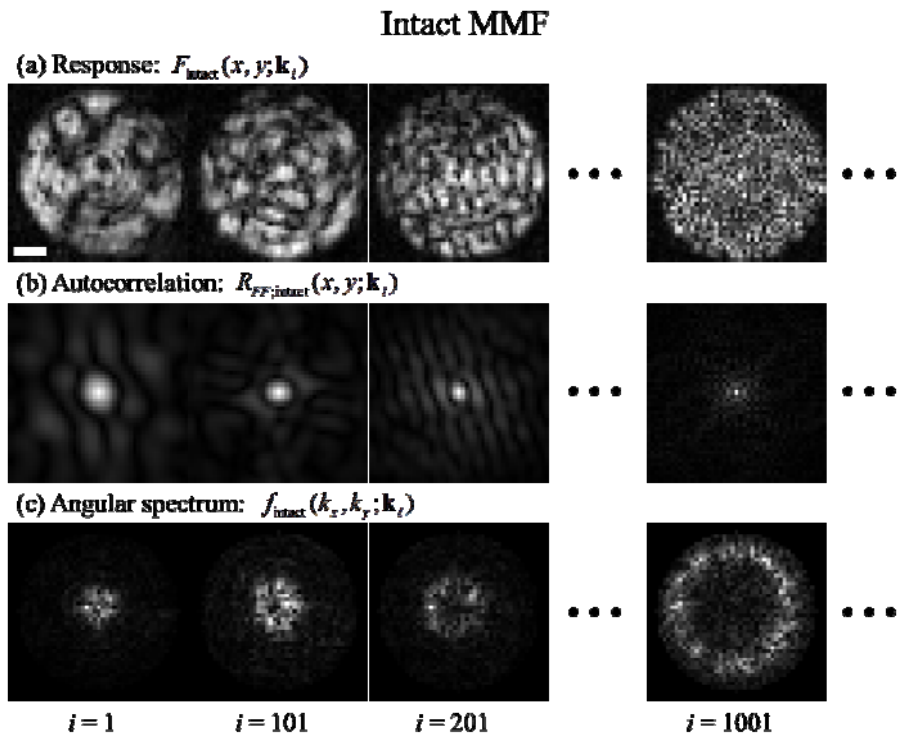

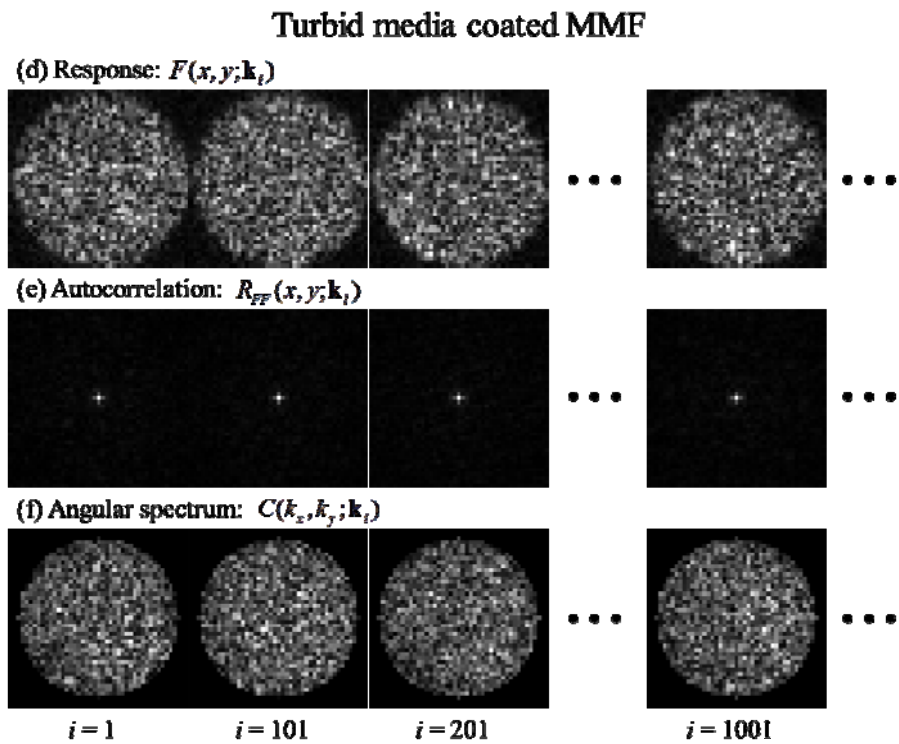

Fig. 3. (a) Recorded responses $F_{\text{intact}}(x, y; \mathbf{k}_i)$, (b) amplitude of the autocorrelation for the response, $R_{FF;\text{intact}}(x, y; \mathbf{k}_i)$, and (c) angular spectrum of the response, $f_{\text{intact}}(\mathbf{k}''; \mathbf{k}_i)$, of intact MMF, and (d) recorded responses $F(x, y; \mathbf{k}_i)$, (e) amplitude of the autocorrelation for the response, $R_{FF}(x, y; \mathbf{k}_i)$, and (f) angular spectrum of the response, $C(\mathbf{k}''; \mathbf{k}_i)$, of coated MMF. Among the responses of all the incident angles covering the NA of 0.22, the 1st, the 101st, the 201st, and the 1001st of them are presented. Only the intensities are shown here. Scale bar: 10 μm.

Now consider the TM of the coated MMF with a turbid medium deposited in the input facet. The response $F(x, y; \mathbf{k})$ of the coated MMF to $P(x, y; \mathbf{k})$ can be obtained in two steps. First, the response $T(x, y; \mathbf{k})$ of the turbid medium placed at the input of the intact MMF to $P(x, y; \mathbf{k})$ is obtained. Second, the response $F(x, y; \mathbf{k})$ of the intact MMF to $T(x, y; \mathbf{k})$ is obtained. $T(x, y; \mathbf{k})$ is available in [Eq. (6)]. $F(x, y; \mathbf{k})$ is derived as

$$F(x, y; \mathbf{k}) \underset{(a)}{=} \sum_{\mathbf{k}' \in \mathcal{K}'} t(\mathbf{k}'; \mathbf{k}) F_{\text{intact}}(x, y; \mathbf{k}')$$
$$\underset{(b)}{=} \sum_{\mathbf{k}'' \in \mathcal{K}''} \sum_{\mathbf{k}' \in \mathcal{K}'} t(\mathbf{k}'; \mathbf{k}) f_{\text{intact}}(\mathbf{k}''; \mathbf{k}') P(x, y; \mathbf{k}'') \quad (16)$$
$$= \sum_{\mathbf{k}'' \in S_{\text{coated}}(\mathbf{k})} C(\mathbf{k}''; \mathbf{k}) P(x, y; \mathbf{k}'')$$

where (a) is from [Eq. (6)] and the fact that $F_{\text{intact}}(x, y; \mathbf{k}')$ is the response wave of the intact MMF to $P(x, y; \mathbf{k}')$, (b) is from [Eq. (10)], and $S_{\text{coated}}(\mathbf{k}) := \{\mathbf{k}'' \mid C(\mathbf{k}''; \mathbf{k}) \neq 0\}$ denotes the set of excited propagation modes in the MMF when plane waves $P(x, y; \mathbf{k})$ with the mode $\mathbf{k}$ are inserted. Here, different from that of the intact MMF, the number of excited modes in the coated MMF for $F(x, y; \mathbf{k})$ is not small (Fig. 3 (f)). We can easily see in [Eq. (13)] that $S_{\text{coated}}(\mathbf{k}) = \bigcup_{\mathbf{k}'} S_{\text{intact}}(\mathbf{k}')$, the set of excited modes in the coated MMF is the union of all the sets of excited modes in the intact MMF. This is because the random scrambling in the turbid medium varies the directions of the waves. This makes the MMF to have incoming waves with a variety of incident angles, and the propagation modes corresponding to those incident angles are all excited. Now, $F(x, y; \mathbf{k})$ are made by combining many $P(x, y; \mathbf{k}'')$ with a variety of $\mathbf{k}''$. Thus, there are many possible incoherent $F(x, y; \mathbf{k})$ patterns. It becomes easier to find many incoherent interference patterns out of them. As a result, it gets easier to compose the TM with many incoherent interference patterns. In Fig. 3 (d) and (f), we see that the responses of the coated MMF to plane waves with the considered incident angles $\theta_z$ are complex enough to be speckle patterns.

Having more propagation modes in the coated MMF surely provides a better situation for the TM to be incoherent. But, this does not always mean that the TM would be incoherent. For an incoherent TM, the way of combining of the propagation modes $C(\mathbf{k}''; \mathbf{k})$ needs to be incoherent for the beams of light with different incident angles $\mathbf{k}$. It would serve no point if the way of combining the propagation modes was the same for all the incident beams considered, then, as the TM would be completely coherent. We can see in [Eq. (13)] that the way of combining becomes different if $t(\mathbf{k}'; \mathbf{k})$ is different from each other for $\mathbf{k}$. For two different modes $\mathbf{k}_i$ and $\mathbf{k}_j$ ($\mathbf{k}_i \neq \mathbf{k}_j$), it is found that the contributions $t(\mathbf{k}'; \mathbf{k}_i)$ and $t(\mathbf{k}'; \mathbf{k}_j)$ are uncorrelated if the angle difference of the two modes is not too small, $\cos^{-1}(\mathbf{k}_i \cdot \mathbf{k}_j) \geq \delta$ for a certain small $\delta$ [26]. Please be reminded that a random matrix generation gives an incoherent TM (Sec. 3.2). Now, with the incoherent combinations of the coated MMF, we expect the TM to be more incoherent than that of the intact MMF.

We compare the incoherence $\mu_3$ ( $0 \leq \mu_3 \leq 1$ ) of the TMs of the intact MMF and the coated MMF in Fig. 4. The respective number of rows and that of columns of TM are 2025 and 2000 for the intact MMF and 2025 and 4000 for the coated MMF. We consider several values of $\alpha$ for $1 \geq \alpha \geq 0$. Please note that it is desirable to have smaller $\mu_3$ for better incoherence. $\mu_3$ of the intact MMF starts to have nonzero value from $\alpha = 0.75$ and becomes larger as $\alpha$ decreases. The start of nonzero value of $\mu_3$ for the coated MMF is at $\alpha = 0.65$ and $\mu_3$ becomes larger as $\alpha$ decreases. For $0.65 \geq \alpha \geq 0.45$, the ratios of the $\mu_3$ of the intact MMF and that of the coated MMF tend to increase (4 at $\alpha = 0.66$ and 6.55 at $\alpha = 0.45$). For $0.4 \geq \alpha \geq 0.1$, the ratios tend to decrease (6.09 at $\alpha = 0.4$ and 4.5 at $\alpha = 0.1$). Regardless of the tendency, the ratios are considerable for $0.65 \geq \alpha \geq 0.1$. At $\alpha = 0.05$, the $\mu_3$ for the coated MMF is larger than that of the intact MMF, and they become the same at $\alpha = 0$. Though the $\mu_3$ for the intact MMF is better at $\alpha = 0.05$, the incoherence here is not meaningful because we cannot say that $\alpha = 0.05$ is comparable to 1. In all the values of $\alpha$ which are reasonably comparable to 1 ( $0.75 \geq \alpha \geq 0.1$ ), it is found that the coated MMF has even better incoherence than the intact MMF does.

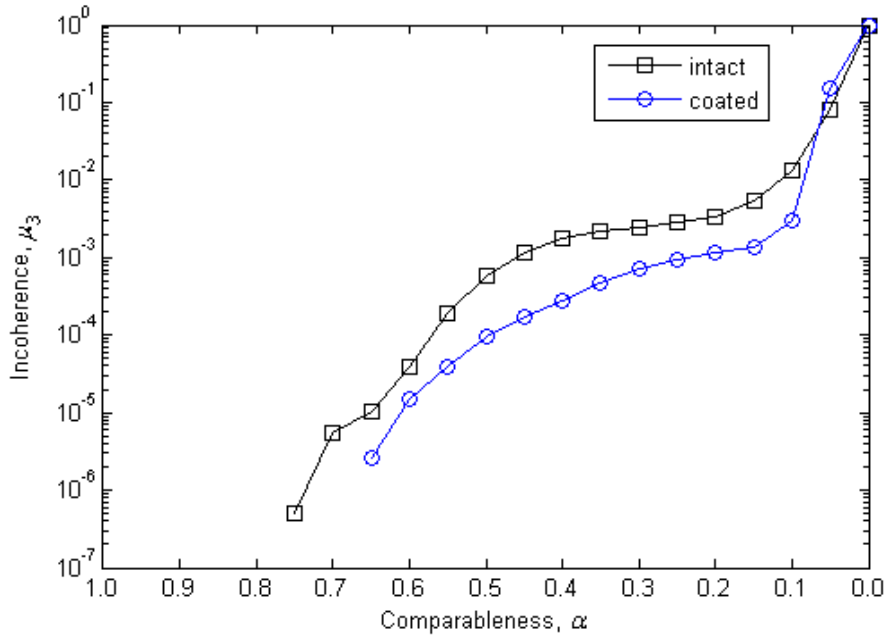

Fig. 4. The incoherence of TMs for intact MMF ($M$ = 2025 and $N$ = 2000) and coated MMF ($M$ = 2025 and $N$ = 4000). The incoherence of them is plotted in log scale.

## 4. Results

We now aim to compare the object image reconstruction capabilities with and without efficiently utilizing the HR encoding effect. For the conventional object wave reconstruction, we use PINV (Section 2.2). For the proposed reconstruction, we use the SR framework which is reported to take advantage of the HR encoding effect in a satisfactory manner [19]. For the sparse recovery in the SR framework, we use the alternating direction method [21] for its efficiency. For the sparsifying basis in the sparse recovery, we use the Fourier basis directly, $\mathbf{\Psi} = \mathbf{I}$. For a fair comparison between the two reconstructions, the reconstructed images are normalized so that their norms become one. For the TM measurements, $N = 2000$ and $N = 4000$ incident angles of the incoming waves are considered and the responses of them are captured respectively for the intact MMF and the coated MMF. The TMs cover the NA of 0.22 ( $0° \leq \theta_z \leq 12.71°$ ) and the NA of 0.4 ( $0° \leq \theta_z \leq 23.58°$ ). The dimension of $M$, the number of pixels in the CCD used in our experiment, corresponding to the output NA of 0.24 (Sec. 2.1) is 2025 ($M = 2025$). Thus, the TMs have the dimensions ($M \times N$) $2025 \times 2000$ and $2025 \times 4000$ for the intact fiber and for the coated fiber, respectively.

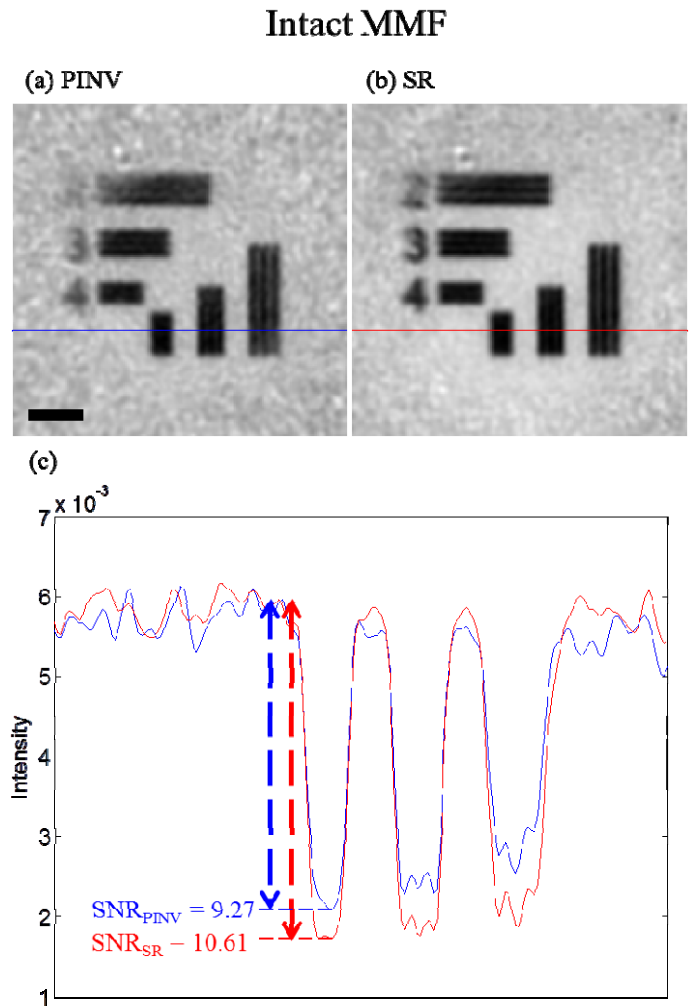

Fig. 5. Reconstruction for imaging through intact MMF. (a) Recovered amplitude image using PINV. (b) Recovered amplitude image using SR. (c) Cross sections of them. Images are averaged over 1000 samples. Scale bar: 10 μm. SNRs are calculated in the cross sections.

Fig. 5 shows the reconstructed images when the image is transferred through the MMF without depositing the turbid medium on the input facet of the fiber, hence there is no HR encoding effect as it should be. It is seen that the smallest structures in the object image cannot be resolved for both reconstructions. This is reasonable since higher modes of object waves which have incident angles beyond the acceptance angle of the intact MMF are not captured. Here, both PINV and SR are shown to reconstruct the given signal well without significant perturbations; the system is not underdetermined. We found some image quality improvements when the SR is employed. For example, pay attention to the quality of reconstruction of the alphanumeric numbers 2, 3, and 4 shown at the image shown in Fig 5. The number 2 at the upper left corner of the image becomes reconstructed and the numbers 3 and 4 becomes clearer. However, the improvement overall by SR is not very significant. The signal-to-noise ratio (SNR) is increased from 9.27 to 10.61; just a 14% increment. Here, the SNR was calculated as SNR $= |\bar{S}_{sig} - \bar{S}_{bg}|/\sigma_{bg}$ where $\bar{S}_{sig}$ is the mean of the signal patterns, $\bar{S}_{bg}$ is the mean of the background, and $\sigma_{bg}$ is the standard deviation of the background.

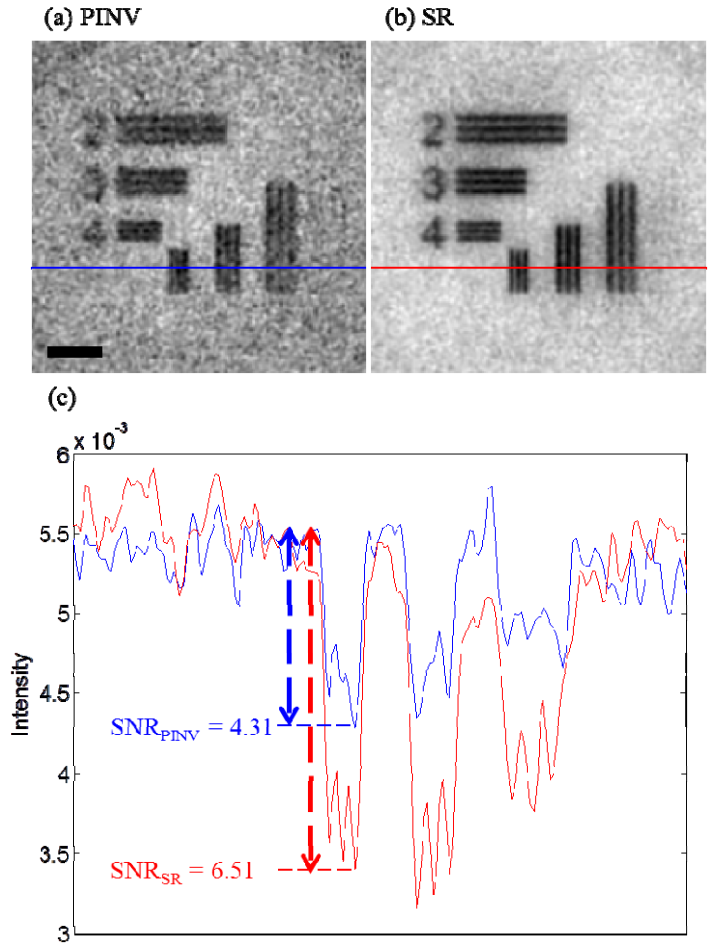

Turbid media coated MMF
(a) PINV  (b) SR
(c)
$SNR_{PINV} = 4.31$
$SNR_{SR} = 6.51$

Fig. 6. Reconstruction for imaging through coated MMF. (a) Recovered amplitude image using PINV. (b) Recovered amplitude image using SR. (c) Cross sections of them. Images are averaged over 2000 samples. Scale bar: 10 μm. SNRs are calculated in the cross sections.

Now, consider the recovered images when the turbid medium is used. It is found in Fig. 6 that the structures in the patterns are significantly improved in terms of resolution. This is true regardless of the use of recovery routines PINV vs. SR. Both methods deliver much improved reconstruction fidelity compared to those of the intact MMF. This is because higher mode waves are introduced through the turbid medium. However, as it was already discussed, the use of turbid medium makes the system underdetermined. This makes the reconstruction of the object image difficult. As a result, it is shown that the reconstructed image by PINV becomes significantly perturbed by speckles. The SNR also becomes reduced considerably compared to that with the intact MMF.

We now show the results employing the SR framework for object wave recovery. As discussed earlier, the point here is to see if it will bring forth improved quality in reconstruction via efficient utilization of the HR encoding process offered by the use of turbid medium. As expected, in contrast to the case of PINV, SR is shown to improve the reconstruction significantly well. The speckle is successfully removed in the reconstructed image. We can also see all the small scale structures in the recovered image. The SNR becomes increased from 4.31 to 6.51; a 51% increment. With these, we have verified that the SR framework can exploit the HR encoding process of the turbid medium and improve the quality of the image reconstruction.

## 5. Conclusion

In conclusion, we demonstrated that the random scattering in the turbid media can be exploited for improving the quality of image reconstruction in MMF imaging. Random scattering through turbid medium provides random encoding of the object signal in holistic and incoherent manner. This encoding can be efficiently utilized in the signal recovery process within the proposed sparse representation framework. As a result, the perturbation is significantly reduced, the image contrast becomes sharper, and the fine details within the image can be captured.

**Acknowledgments**

This work was supported by the National Research Foundation of Korea (NRF) grant funded by the Korean government (MEST) (Do-Yak Research Program, No. 2013-035295).